\documentclass[runningheads]{svmult}

\usepackage{makeidx}   
\usepackage{graphicx}  
\usepackage{subeqnar}  
\usepackage{multicol}  
\usepackage{physprbb}  
\makeindex             

\begin{document}
\title*{Constraining Dark Matter with the Long-Term Variability of Quasars}
\toctitle{Constraining Dark matter with
\protect\newline the Long-Term Variability of Quasars}
%
%
\titlerunning{Constraining Dark Matter with the Long-Term Variability of Quasars}
%
\author{Erik Zackrisson\inst{1}
\and Nils Bergvall\inst{1}
\and Phillip Helbig\inst{2}}
\authorrunning{Erik Zackrisson et al.}
%
%
\institute{Department of Astronomy and Space Physics, Box 515, SE-75120 Uppsala, Sweden
\and Institut f\"ur theoretische Physik, Johann Wolfgang Goethe-Universit\"at\\
Postfach 11 19 32, 60054 Frankfurt am Main, Germany}

\maketitle              

\begin{abstract}
By comparing the results of numerical microlensing simulations to the observed long-term variability of quasars, strong upper limits on the cosmological density of compact objects in the mass range $10^{-2}$--$10^{-4} \ M_\odot$ may be imposed. Using recently developed methods to better approximate the amplification of large sources, we investigate in what way the constraints are affected by assumptions concerning the size of the optical continuum-emitting region of quasars in the currently favored ($\Omega_\mathrm{M}=0.3$, $\Omega_\Lambda=0.7$) cosmology.
\end{abstract}

\section{Introduction} 
Although the optical variability of quasars on time scales of a few years could be both intrinsic in nature (e.g. due to accretion disk instabilities or supernova explosions) and caused by microlensing along the line of sight, adding the effects of both mechanisms can only increase the probability of variations. By assuming that {\it all} variability is due to microlensing and comparing the predictions from microlensing scenarios to the observed amplitudes of quasar light curves, upper limits to the cosmological density of compact object may therefore be imposed. This technique was first implemented in \cite{zackrissoneref1} to constrain compact dark matter populations in the mass range 1--10$^{-4} \ M_\odot$ for an Einstein--de Sitter universe. Here, we extend the analysis to the $\Omega_\mathrm{M}=0.3$, $\Omega_\Lambda=0.7$ cosmology and the full range of quasar sizes allowed.
 
\section{Analysis}
In our model of quasar microlensing, the multiplicative magnification assumption outlined in \cite{zackrissoneref2} is assumed to adequately reproduce the statistical probability of variability. The results from the numerical simulations are compared to the observational sample of \cite{zackrissoneref3}, containing 117 quasars in the redshift range $z=0.29$--$3.23$, monitored for ten years. The method of comparing observed and synthetic samples closely follows that of \cite{zackrissoneref1}, which may be consulted for further details. 

The constraints derived assume that the luminosities of these objects are independent of the radii of their optical continuum-emitting regions, $R_\mathrm{QSO}$. Under this condition, the most conservative upper limits are inferred from volume-limited synthetic samples. Our analysis is therefore restricted to this case only. 

\section{Upper Limits on The Cosmological Density of Compact Objects}
The size of the optical continuum-emitting region of quasars is not a well-determined quantity, but typically assumed to lie somewhere in the range $R_\mathrm{QSO}=10^{12}$--$10^{14}$ m. Whereas only the $R_\mathrm{QSO}=10^{13}$ m case was considered when deriving the constraints in \cite{zackrissoneref1}, we here investigate the impact that smaller and larger quasar sizes may have on the upper limits inferred.

\begin{figure}[h]
\begin{center}
\includegraphics[scale=0.33]{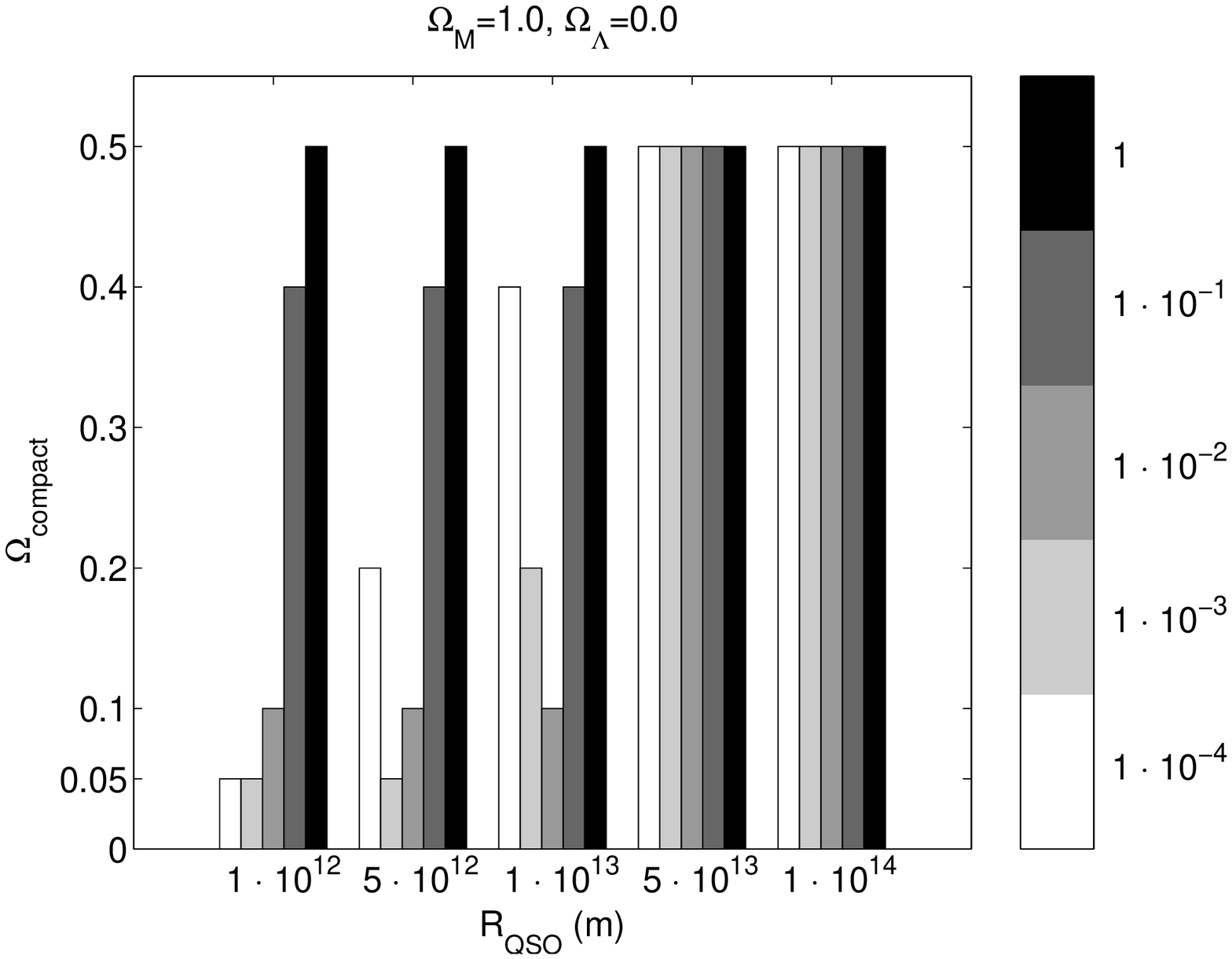}
\includegraphics[scale=0.33]{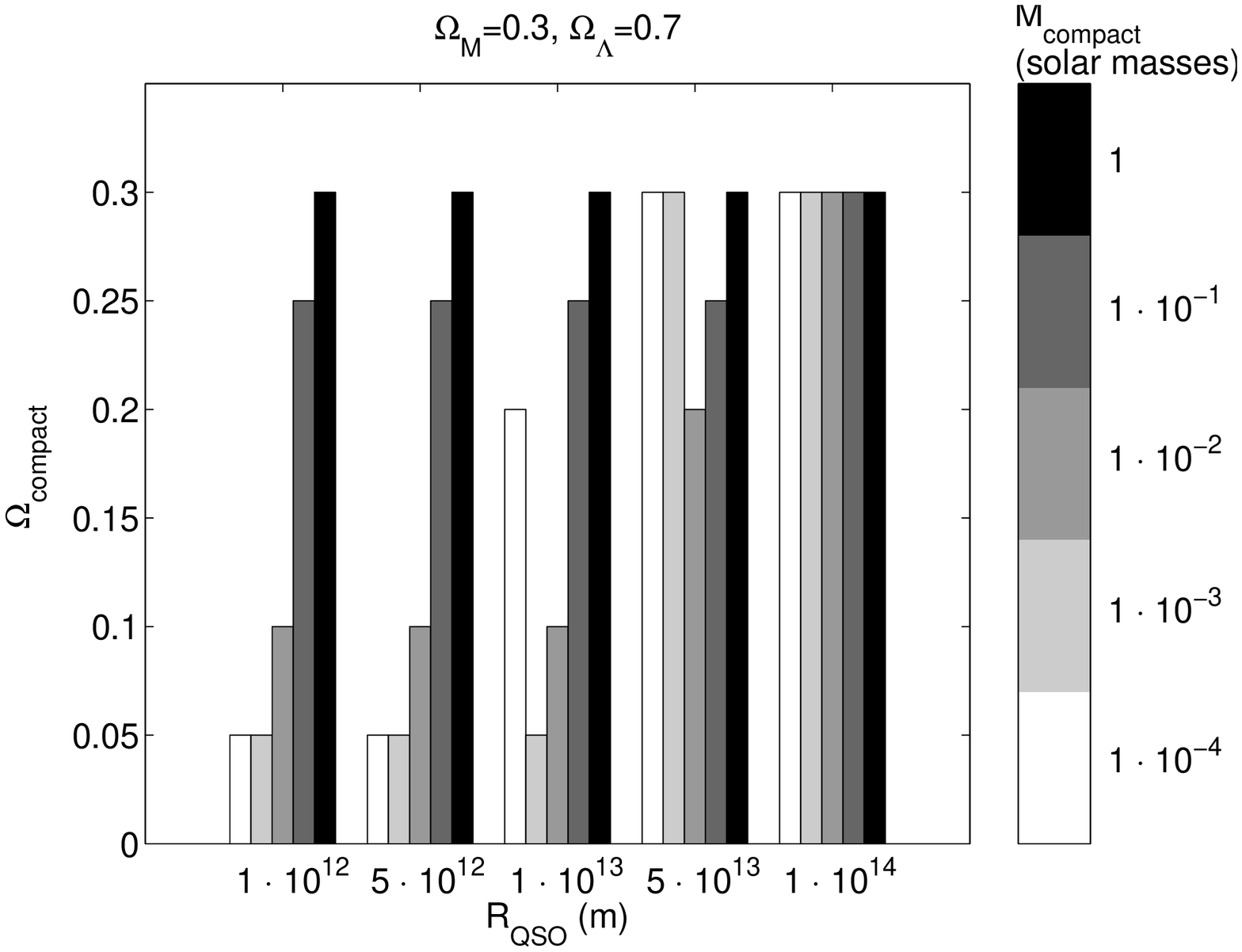}
\end{center}
\caption[]{Allowed regions of the ($M_\mathrm{compact}$, $R_\mathrm{QSO}$, $\Omega_\mathrm{compact}$) parameter space in the Einstein--de Sitter (\textbf{Left}) and $\Omega_\mathrm{M}=0.3$, $\Omega_\Lambda=0.7$ (\textbf{Right}) cosmologies. Lower values of $\Omega_\mathrm{compact}$ indicate stronger constraints}
\label{zackrissonefig1}
\end{figure}

Figure \ref{zackrissonefig1} indicates the constraints on the cosmological density of compact objects, $\Omega_\mathrm{compact}$, for different lens masses, $M_\mathrm{compact}$, and $R_\mathrm{QSO}$ in both Einstein-de Sitter and $\Omega_\mathrm{M}$=0.3, $\Omega_\Lambda$=0.7 cosmologies.  Poisson probabilities of 10\% (see \cite{zackrissoneref1}) have been chosen to define the upper limits. A Hubble constant of $H_0=65$ km/s/Mpc has also been assumed. 

Even though the transition to the $\Lambda$-dominated universe significantly strengthens the constraints, the limits on $\Omega_\mathrm{compact}$ are seen to be very sensitive to the value of $R_\mathrm{QSO}$ used. 

\section{The Impact of Large-Source Microlensing}
\begin{figure}[h]
\begin{center}
\includegraphics[scale=0.4]{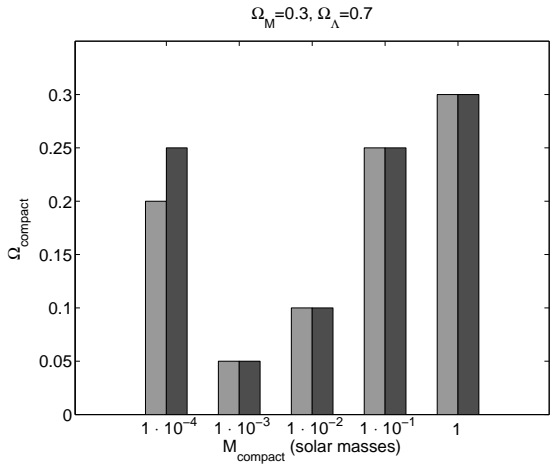}
\end{center}
\caption[]{The allowed region of the ($M_\mathrm{compact}$, $\Omega_\mathrm{compact}$) parameter space for a source size of $R_\mathrm{QSO}=1\cdot 10^{13}$ m in the $\Omega_\mathrm{M}=0.3$, $\Omega_\Lambda=0.7$ cosmology, with (\textit{light gray}) and without (\textit{dark gray}) the use of the amplification formula for large sources (\cite{zackrissoneref4})}
\label{zackrissonefig2}
\end{figure}

In \cite{zackrissoneref4}, the amplification formula derived in \cite{zackrissoneref1} was shown to underpredict the amplification of large sources and replaced by a modified expression. As predicted in \cite{zackrissoneref4}, the improved amplification formula should therefore result in tighter limits on $\Omega_\mathrm{compact}$. In Fig.~\ref{zackrissonefig2}, however, we show that  the implementation of the large-source formalism merely has a modest effect on the upper limits, and only alters the constraints at the lowest mass ($10^{-4}$ $M_\odot$) considered.

\section{Conclusions}
By comparing the results of numerical microlensing simulations to the observed long-term variability of quasars, strong upper limits may be placed on the cosmological density of dark matter in the form of compact objects. In the currently favored $\Lambda$-dominated universe ($\Omega_\mathrm{M}=0.3$, $\Omega_\Lambda=0.7$), compact objects with masses $\sim10^{-2}$ $M_\odot$ and $\sim10^{-3}$ $M_\odot$ cannot contribute more to the cosmological density than $\Omega_\mathrm{compact} = 0.1$ and $\Omega_\mathrm{compact} = 0.05$ respectively,  provided that the continuum-emitting region of quasars is not significantly larger than $1\cdot10^{13}$ m. We do however note that the method used is sensitive to the assumptions concerning the size of the continuum-emitting region of quasars, and becomes effectively useless for $R_\mathrm{QSO} \ge 5\cdot10^{13}$ m.

The large source amplification formula developed in \cite{zackrissoneref4} is shown not to have any significant impact in the part of parameter space for which the long-term variability method provides the most interesting constraints.

A more detailed analysis of the uncertainties present in this technique to constrain the cosmological density of compact objects is currently underway (\cite{zackrissoneref5}).

%


\begin{thebibliography}{8.}
\addcontentsline{toc}{section}{References}

\bibitem{zackrissoneref1}
P. Schneider: A\&A \textbf{279}, 1 (1993)

\bibitem{zackrissoneref2} 
J.P. Ostriker, M. Vietri: ApJ \textbf{267}, 488 (1983)

\bibitem{zackrissoneref3} 
M.R.S. Hawkins, P. V\'eron: MNRAS \textbf{260}, 202 (1993)

\bibitem{zackrissoneref4}
G. Surpi, S. Refsdal, P. Helbig: submitted to A\&A

\bibitem{zackrissoneref5}
E. Zackrisson, N. Bergvall: in preparation

\end{thebibliography}
\end{document}